\documentstyle[times,psfig,graphics,astrobib,amssymb]{basi}

\newcommand{\be}{\begin{equation}}
\newcommand{\e}{\end{equation}}
\newcommand{\bear}{\begin{eqnarray}}
\newcommand{\ear}{\end{eqnarray}}

\begin{document}
\title{Physics of Structure Formation in the Universe}
\author[T. Roy Choudhury]{T. Roy Choudhury
\thanks{e-mail:tirth@iucaa.ernet.in}\\
Inter University Centre for Astronomy and Astrophysics, Pune - 411 007}
\maketitle
\begin{abstract}
In recent years, unprecedented progress in observational cosmology  has
revealed a great deal of information about the  formation and evolution of
structures in the universe. This,  in turn, has raised many challenging issues 
for the theorists. In the thesis,  we have addressed  two such key issues,
namely,  (a) the formation of  baryonic structures and (b) the nature of dark
matter and dark energy and  the limitations in determining their nature from
observations. The main results from the thesis are:
(i) The baryons in the intergalactic medium at redshifts $z \sim 2.5$ can  be
modelled (both analytically and  semi-analytically) by accounting for the 
non-linearities in the density field through lognormal approximation.  Our
results  agree with observations, and can be used for constraining  parameters
related to the baryons.
(ii) A simple model based on baryon conservation, along with  observational
estimates of cosmic star formation rate,  correctly predicts the abundance of
damped Lyman-$\alpha$ systems  in the universe.
(iii) The  redshift distribution of gamma ray bursts can, in principle, be
used  for studying the physical conditions of the universe before  reionization
epoch (which is otherwise a difficult task). 
(iv) It might be possible that the dark matter and dark energy arise from  the
same scalar field, provided the equation of state has a  dependence on the
length scale. The possibility of using the  kinematical and geometrical
measurements (such as supernova observations) for determining  the nature and
evolution of the dark energy is discussed.
\end{abstract}


\section{Introduction}
\vspace{-0.5cm}

In recent years, our understanding of the universe has been driven by
tremendous progress in observational cosmology  which has revealed 
a great deal of 
information about the geometry, mass distribution and composition, and
formation and evolution of structures of the universe. 
The high accuracies in the 
measurement of most of the cosmological parameters helped 
the theoretical cosmologists to develop the so called 
``standard model'' of cosmology -- a model which is 
consistent with all the observations.
However, these 
observations, have also left the theoretical 
cosmologists with some challenging issues. 
For example, we do not have any laboratory evidence for about 
95 per cent of the matter component in the universe. 
About one third of these is a non-relativistic pressure-less fluid 
which does not interact with radiation, while the rest 
two-third is a negative pressure component. 
Similarly, we still do not have 
satisfactory models which describe the physics of the 
formation of baryonic structures, like galaxies, in the universe. 
In this thesis, we have addressed two key
issues in this field, namely, (i) the formation of baryonic structures 
(Sections 2--5) and
(ii) the nature of dark matter and dark energy and 
the limitations in determining their nature from observations 
(Section 6). 
The analytical and 
semi-analytical models described in this thesis 
are found to match excellently with 
current observational data, and can be further compared with 
improved datasets expected in near future.

\section{Analytical modelling of the neutral hydrogen distribution 
in the universe}
\vspace{-0.5cm}

A significant fraction of baryons at $z \le 5$ is found in the 
form of a diffuse ionised intergalactic 
medium (IGM), which is usually probed through the absorption 
spectra of distant QSOs. Most of the low neutral
hydrogen column density absorption lines (commonly called 
`Ly$\alpha$' clouds) in a typical QSO spectrum 
are believed to be due to low-amplitude baryonic
fluctuations in the IGM. 
In this section, we model the distribution of neutral hydrogen 
in the universe at redshifts 2--4 and compare with 
observations of QSO spectra. 
We use an approximation scheme for the 
non-linear baryonic mass density -- the lognormal ansatz. 
This ansatz
was used earlier by \citeN{bd97} to perform one-dimensional
simulations of lines of sight and analyse the properties of
absorption systems. We have taken a completely analytical approach, which
allows us to explore a wide region of the parameter space for our model. 
We have also assumed that 
the gas in the IGM is in a state of photoionization equilibrium, with 
the temperature and density of the gas being related 
by a simple power-law (usually called the equation of state of the IGM).
The analytical results have been compared with 
observations to constrain various cosmological 
and IGM parameters, whenever possible \cite{cps01}. 

\begin{figure}
\begin{center}
\psfig{figure=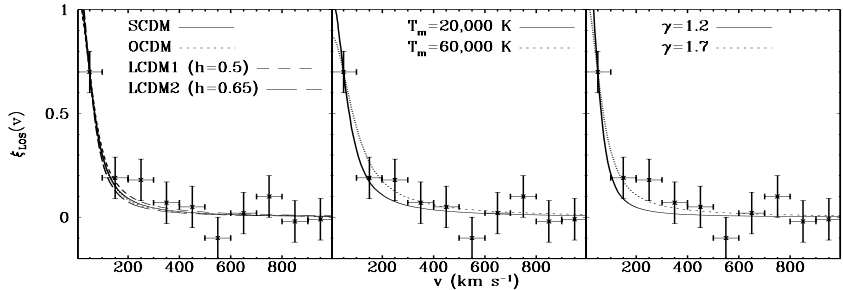,width=0.9\textwidth}
\caption{
Comparison of the theoretical $\xi_{\rm LOS}$ with observational data 
(Cristiani et al. 1997). The
theoretical curves have been normalized in such a way that they match with the
observed data point at the lowest velocity bin. 
We show the dependence of the LOS correlation on 
background cosmological model (left), 
Jeans temperature $T_m$ (centre) and slope of the equation of state 
$\gamma$ (right).
It is clear that one cannot constrain the background cosmological
model without constraining the IGM parameters, especially $\gamma$.
}
\label{fig:los}
\end{center}
\end{figure}

\begin{figure}
\begin{center}
\psfig{figure=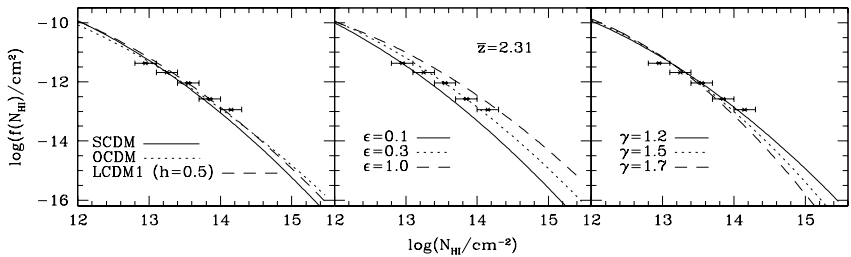,width=0.9\textwidth}
\caption
{
The column density distribution for neutral hydrogen $f(N_{\rm H{\sc i}})$ 
for redshift 2.31. The data points with error-bars 
are obtained from Hu et al. (1995) 
and Kim et al. (1997) 
We show the dependence of $f(N_{\rm H{\sc i}})$ on 
background cosmological model (left), 
parameter $\epsilon$ relating the column density to the density of the gas 
(centre) and slope of the equation of state 
$\gamma$ (right).
}
\label{fig:nh}
\end{center}
\end{figure}

The analytical predictions for the line-of-sight (LOS) correlation of 
the neutral hydrogen distribution is compared with observational data 
obtained from the absorption spectra of distant QSOs in Figure \ref{fig:los}.
\nacite{cdd++97}
We find that the effects on the LOS correlation 
owing to changes in cosmology and the slope of the 
equation of state of the IGM, $\gamma$ are 
of the same order, which means that we cannot constrain both the 
parameters simultaneously. 
Our models also reproduce the observed column density distribution 
for neutral hydrogen, and the shape of 
the distribution depends on $\gamma$ (see Figure \ref{fig:nh}). 
\nacite{hkcsr95} 
\nacite{khcs97}
It is evident from the figure that one can rule out $\gamma > 1.6$ for 
$z \simeq 2.31$ using 
the column density distribution.

\section{Absorption systems in the universe: Comparison of semi-analytical 
model with observations}

\vspace{-0.5cm}

\begin{figure}
\begin{center}
\psfig{figure=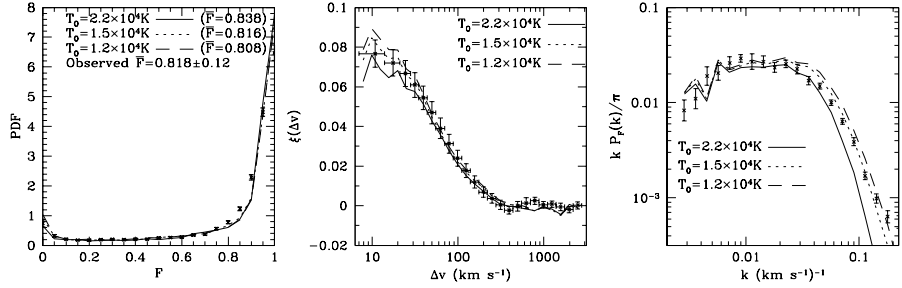,width=\textwidth}
\caption
{
Comparison between simulations and observed results for 
$f \equiv (\Omega_B h^2)^2/J_{-12} = 0.026^2$,  
$\gamma = 1.5$ and three values of $T_0$ as indicated in the 
figure. The 
points with error-bars are the observed data points (McDonald et al. 2000). 
We have plotted the probability distribution function (left), the 
correlation function (middle) and the power spectrum (right) of the 
transmitted flux.
}
\label{fig:stat_allowed}
\end{center}
\end{figure}

We extend the analytical calculations of the previous section 
and perform one-dimensional semi-analytical 
simulations along the lines of sight to model the IGM. 
Since this procedure is computationally 
efficient in probing the parameter space -- and reasonably accurate -- we 
use it to recover the values of various parameters related to the IGM 
(for a fixed background cosmology) 
by comparing the model predictions with different observations 
\cite{csp01}.
For the $\Lambda$ cold dark matter model  
($\Omega_m=0.4$, $\Omega_{\Lambda}=0.6$ and $h=0.65$), 
we find that the statistics obtained from the transmitted flux 
of the simulated absorption spectrum of QSOs match excellently 
with observations (see Figure \ref{fig:stat_allowed}) 
at a mean redshift $z \simeq 2.5$.

Further, using these 
transmitted flux statistics, we obtain  
constraints on (i) the combination $f=(\Omega_B h^2)^2/J_{-12}$, where 
$\Omega_B$ is the baryonic density parameter and $J_{-12}$ is the 
total photoionization rate in units of $10^{-12}$s$^{-1}$, 
(ii) 
temperature $T_0$ corresponding to the mean density, and 
(iii) the slope $\gamma$ of the effective 
equation of state of the IGM.
\nacite{mmr++00}
\begin{figure}
\begin{center}
\psfig{figure=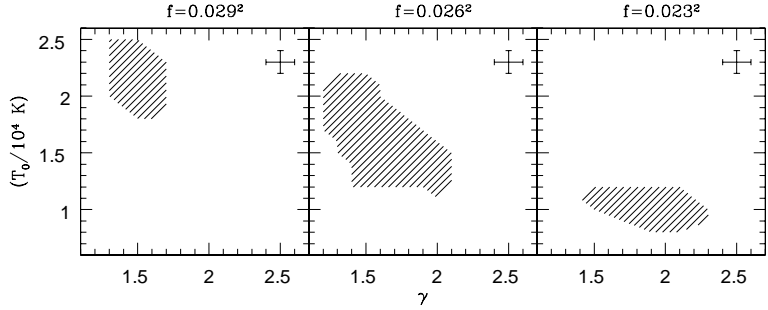,width=0.8\textwidth}
\caption
{
Constraints obtained in the $\gamma-T_0$ space for different values of 
$f$, {\it using transmitted flux statistics}. The shaded regions 
denote the range allowed by observations. The boundaries are 
uncertain by an amount 0.1 along $\gamma$ axis and by 1000K along the 
$T_0$ axis because of finite sampling, which is shown by a cross at the upper
right hand corners of the panels.
}
\label{fig:t0_gamma}
\end{center}
\end{figure}
As an example, we show 
the constrained parameter space in the 
$\gamma - T_0$ plane for three values 
of $f$ in Figure \ref{fig:t0_gamma}. 
It is obvious that as we go to lower values of $f$, the observations allow 
lower values of $T_0$ and higher values of $\gamma$. 
In general. 
we find that 0.8 $<T_0/(10^4 {\rm K})<$ 2.5 and $1.3<\gamma<2.3$, 
while the constraint obtained on $f$ is $0.020^2<f<0.032^2$. 
A reliable lower bound on $J_{-12}$ can be used to put a lower bound on 
$\Omega_B h^2$, which
can be compared with similar constraints obtained from big bang 
nucleosynthesis (BBN) and cosmic microwave background radiation (CMBR) 
studies. 

\section{Abundance of damped Ly$\alpha$ absorbers: A simple analytical model}
\begin{figure}
\begin{center}
\psfig{figure=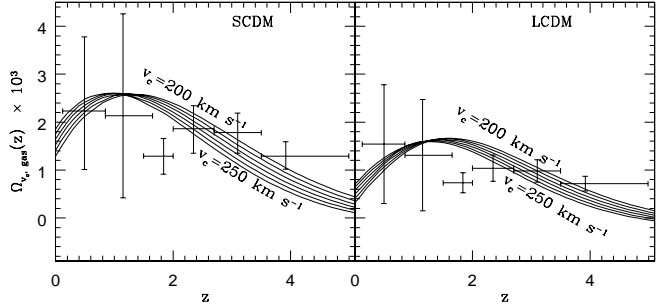,width=0.7\textwidth}
\caption
{Density of matter contained in gaseous form $\Omega_{v_c, {\rm gas}}(z)$ 
as a function of 
$z$, plotted for two cosmological models. The circular velocity ranges 
from  $v_c=200$ km s$^{-1}$ to $v_c=250$ km s$^{-1}$ in both plots. 
The observed data points with error-bars are
obtained from Peroux et al. (2001). 
As can be seen from the figure, the ballpark estimate of 
$\Omega_{v_c,{\rm gas}}(z)$ from our simple model is well within 
observational constraints.
}
\label{fig:omega_gas_6vc}
\end{center}
\end{figure}

\vspace{-0.5cm}
This section deals with comparatively high density regions, namely, the 
damped Ly$\alpha$ systems (DLAs).
The DLAs are identified with 
the lines having highest 
column densities in a typical observed absorption spectrum of a 
distant quasar.
These high 
column density systems are important in understanding 
the baryonic structure formation, because they 
contain a fair amount of the neutral hydrogen in the universe at 
high redshifts.
As a preliminary study for 
understanding these systems, a simple analytical model for 
estimating the fraction ($\Omega_{\rm gas}$) 
of matter  
in gaseous form within the collapsed dark matter (DM) haloes is presented. 
The  model is developed using  (i) the Press-Schechter formalism 
to estimate the fraction of baryons in DM haloes 
and (ii) the observational estimates 
of the star formation rate at different redshifts. The prediction for 
$\Omega_{\rm gas}$ from the model is in broad agreement with 
the observed abundance of the damped Ly$\alpha$ systems, which 
is shown in Figure \ref{fig:omega_gas_6vc}. 
\nacite{pmsi01}
Furthermore, it can be used for estimating the 
circular velocities of the collapsed haloes at different redshifts, which 
could be compared with future observations \cite{cp02}.

\section{Probing the reionization epoch with redshift distribution of 
gamma-ray bursts}
\begin{figure}
\begin{center}
\psfig{figure=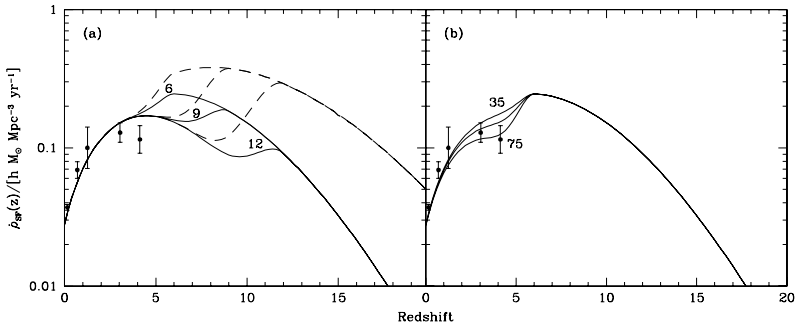,width=0.8\textwidth}
\caption
{The SFR density as a function of
redshift for three epochs of reionization and 
various model parameters [see Choudhury \& Srianand (2002) for details]. 
The curves are normalized using 
the extinction corrected 
data points 
from Somerville, Primack, \& Faber (2001).}
\label{fig:sfr}
\end{center}
\end{figure}

\vspace{-0.5cm}
In this section, we take our first step towards 
modelling the reionization.
We explore the possibility of using the properties of 
gamma-ray bursts (GRBs) to probe the physical conditions in the epochs prior
to reionization. 
The GRBs are among the brightest sources in the sky in any wavelength region.
They are believed to originate due to collapse of very massive stars 
in the galaxies. In this section, 
the redshift distribution of GRBs is modelled 
using the Press-Schechter formalism with an 
assumption that they follow the cosmic
star formation history. 
We reproduce the observed star formation rate (SFR) 
obtained from 
galaxies in the redshift range $0 < z < 5$ (Figure \ref{fig:sfr}), 
as well as the redshift distribution 
of the GRBs inferred from the luminosity-variability 
correlation of the burst light curve \cite{cs02}. 
Interestingly, we find that the fraction of GRBs at high redshifts,  
the afterglows of which cannot be observed in 
$R$ and $I$ band
owing to H{\sc i} Gunn-Peterson optical depth can, at the most,  
account for \emph {one third of the dark GRBs}. This means that a 
substantial fraction of optically dark GRBs ($\gtrsim 66$ per cent) 
originate because of effects such as dust extinction.
The observed redshift distribution of GRBs, 
with much less scatter than the one 
available today, can put stringent constraints on the epoch 
of reionization and the nature of gas cooling in the epochs prior 
to reionization.

\nacite{spf01}
\nacite{cs02}
\section{A model for dark matter and dark energy}

\begin{figure}
\begin{center}
\rotatebox{270}{\resizebox{0.4\textwidth}{!}
{\includegraphics{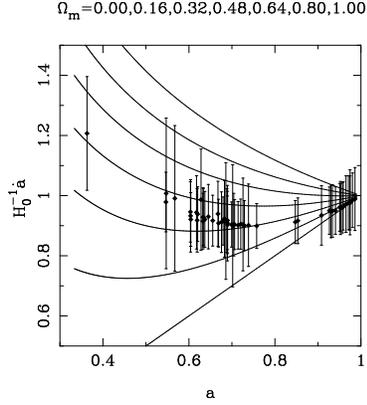}}}
\caption
{The observed supernova data points in the $\dot{a} - a$ plane for 
flat models, where $a$ is the scale factor. 
The procedure for obtaining the data points and the 
corresponding error-bars are described in Padmanabhan \& Choudhury (2002). 
The solid
curves, from bottom to top,  
are for flat cosmological models with 
$\Omega_m = 0.00, 0.16, 0.32, 0.48, 0.64, 0.80, 1.00$ respectively.
}
\label{adotomegam}
\end{center}
\end{figure}
\nacite{pc02b}
This section deals with issues related to the 
nature of dark matter and dark energy.
Current cosmological observations strongly suggest 
the existence of two different kinds of
energy densities dominating at small ($ \lesssim 500$ Mpc) and large ($\gtrsim
1000 $ Mpc) scales. The dark matter component, which dominates at small scales,
contributes $\Omega_m \approx 0.35$ and has an equation of state $p=0$, while
the dark energy component, which dominates at large scales,  contributes
$\Omega_{\Lambda} \approx 0.65$ and has an equation of state $p\simeq -\rho$.
The most obvious
candidate for this dark energy is the cosmological constant (with 
the equation of state $w_X = p/\rho = -1$), which, however, 
raises several theoretical difficulties. This has led to 
models for  dark energy component 
which evolves with time. We 
discuss certain questions related to the 
determination of the nature of 
dark energy component from observations of high redshift supernova.
The main results of our analysis are: 

(i) It is
usual to postulate weakly interacting massive particles (WIMPs) 
for the dark matter component and some form of scalar field
or cosmological constant for the dark energy component. 
We explore the possibility
of a scalar field with a Lagrangian $L =- V(\phi) \sqrt{1 - \partial^i 
\phi \partial_i
\phi}$ acting as {\it both} clustered dark 
matter and smoother  dark energy and
having a scale-dependent equation of state \cite{pc02}.  
This model predicts a relation
between the ratio  $ r = \rho_{\Lambda}/\rho_{\rm DM}$ 
of the  energy densities of the
two dark components and expansion rate $n$ of the universe 
[with $a(t) \propto
t^n$] in the form  $n = (2/3) (1+r) $. For $r \approx 2$, we get $n \approx 2$
which is consistent with supernova observations.

(ii) Although 
the full data set of supernova observations (which are currently available) 
strongly rule out models without dark energy, 
the high ($z > 0.25$) and low ($z < 0.25$) redshift 
data sets, individually, admit decelerating models with 
zero dark energy (see Figure \ref{adotomegam}). 
Any possible evolution 
in the absolute magnitude of the supernovae, if detected, 
might allow the decelerating models to be consistent with 
the data. 

(iii) By studying the sensitivity of the luminosity distance on 
$w_X$, it is possible to argue that although 
one can  determine the present value of $w_X$ accurately from 
the data, one cannot constrain the evolution of $w_X$ \cite{pc02b}.

\vspace{-0.5cm}
\section{Summary}
\vspace{-0.5cm}
The first part of the thesis is devoted to understanding the 
baryonic structure formation in the universe.
We have developed various semi-analytical models to understand 
different aspects of the thermal history of baryons. The models, 
based on some reasonable approximations,
are found to agree with current observations.
In the next part of the thesis, we have addressed a more general 
issue of the nature of dark matter and dark energy in the universe.
Our understanding of the subject is far from complete -- 
the unresolved issues will be settled only through 
future observations. 
There is no doubt that computer simulations will play a major role 
in order to understand such observations -- however the limitations 
due to computational power will always be a hurdle. Thus one needs 
to carry out calculations based on simple ideas and approximations so as 
to keep pace with progress in observational 
cosmology -- like what is done in the thesis.

\vspace{-0.5cm}
\section*{Acknowledgements}
\vspace{-0.5cm}
The work for the thesis was carried out under the supervision of 
T. Padmanabhan. Considerable amount of the work was done in collaboration 
with R. Srianand. 
The author would like to thank the 
Inter-University Centre for Astronomy and Astrophysics, 
Pune, for providing the facilities required to complete the thesis. 
The projects in the initial part of the thesis were supported by the 
Indo-French Centre for 
Promotion of Advanced  Research under contract No. 1710-1.
The author is supported by the 
University Grants Commission, India.

\bibliography{mnrasmnemonic,astropap}
\bibliographystyle{mnras}

\end{document}